\documentclass[12pt]{article}
\usepackage{amssymb,amsmath}
\usepackage{epsfig,psfrag}
\usepackage{curves}

\catcode`\@=11
\def \Tr {\mathop{\rm Tr}\nolimits}
\def \tr {\mathop{\rm tr}\nolimits}

\newcommand\lr[1]{{\left({#1}\right)}}

\newcommand\re[1]{(\ref{#1})}

\newcommand{\ba}{\begin{align}}
\renewcommand{\a}{\alpha}
\renewcommand{\b}{\beta}
\renewcommand{\l}{\lambda}
\newcommand{\dl}{\tilde\lambda}

\renewcommand{\a}{\alpha}

\newcommand{\da}{{\dot\alpha}}
\newcommand{\db}{{\dot\beta}}

\renewcommand{\b}{\beta}

\newcommand{\q}{\theta}

\newcommand{\ep}{\epsilon}

\newcommand{\cN}{{\cal N}}

\newcommand{\p}[1]{(\ref{#1})}
\newcommand{\bt}[1]{{\bar t}}

\newcommand \vev [1] {\langle{#1}\rangle}

\newcommand {\cO}{{{\cal O}}}

\newcommand{\beq}{\begin{equation}}
\newcommand{\eeq}{\end{equation}}
\newcommand{\be}{\begin{equation}}
\newcommand{\ee}{\end{equation}}
\newcommand{\bea}{\begin{eqnarray}}
\newcommand{\eea}{\end{eqnarray}}
\newcommand{\ena}{\end{eqnarray}}
\newcommand{\bear}{\begin{eqnarray}}
\newcommand{\ear}{\end{eqnarray}\noindent}
\newcommand{\la}{\langle}
\newcommand {\ra}{\rangle}

\newcommand{\ft}[2]{{\textstyle\frac{#1}{#2}}}

\def\numberbysection{\@addtoreset{equation}{section}
                     \def\theequation{\thesection.\arabic{equation}}}

\begin{document}


\thispagestyle{empty}

\null\vskip-12pt \hfill
\begin{minipage}[t]{35mm}
 DCPT-10/43 \\
  IPhT--T10/139 \\
 LAPTH--039/10
\end{minipage}

\vskip3.2truecm
\begin{center}
\vskip 0.2truecm {\Large\bf
More on the duality  correlators/amplitudes}

\vskip 1truecm

{\bf    Burkhard Eden$^{*}$, Gregory P. Korchemsky$^{\dagger}$, Emery Sokatchev$^{\ddagger}$ \\
}

\vskip 0.4truecm
$^{*}$ {\it  Durham University, 
Science Laboratories,
 \\
South Rd, Durham DH1 3LE,
United Kingdom \\
 \vskip .2truecm
$^{\dagger}$ Institut de Physique Th\'eorique\,\footnote{Unit\'e de Recherche Associ\'ee au CNRS URA 2306},
CEA Saclay, \\
91191 Gif-sur-Yvette Cedex, France\\
\vskip .2truecm $^{\ddagger}$ LAPTH\,\footnote[2]{Laboratoire d'Annecy-le-Vieux de Physique Th\'{e}orique, UMR 5108},   Universit\'{e} de Savoie, CNRS, \\
B.P. 110,  F-74941 Annecy-le-Vieux, France
                       } \\
\end{center}

\vskip 1truecm 
\centerline{\bf Abstract} 
\medskip
\noindent

We continue the study of $n-$point correlation functions of half-BPS protected operators in $\cN=4$ super-Yang-Mills theory, in the limit where the positions of the adjacent operators become light-like separated. We compute the $l-$loop corrections by making $l$ Lagrangian insertions. We argue that there exists a simple relation between  the
$(n+l)-$point Born-level correlator with
$l$ Lagrangian insertions and the integrand of the $n-$particle
$l-$loop MHV scattering amplitude,
as obtained by the recent momentum twistor  construction of Arkani-Hamed et al. We present several examples of this new duality, at one and two loops.

\newpage

\thispagestyle{empty}


\newpage
\setcounter{page}{1}\setcounter{footnote}{0}



\section{Introduction}

 In this note we present convincing evidence for a new, surprising duality between two apparently different approaches to MHV scattering amplitudes in $\cN=4$ super-Yang-Mills theory. One of them, recently proposed in \cite{us}, consists in treating the {\it integrand} of the amplitude as a four-dimensional conformal object, the  {\it Born-level} correlation function of protected gauge-invariant composite operators with Lagrangian insertions. The other very recent approach \cite{them,Boels:2010nw} also focuses on the {\it integrand} of the amplitude,\footnote{The approach of \cite{them} covers not only MHV, but also all other kinds of amplitudes. Our approach is for the time being restricted to MHV amplitudes only.} regarding it as a four-dimensional conformal object formulated in terms of momentum twistors, and obtained from recursion relations of the BCFW type. Here we demonstrate the exact matching of the two objects for any number of particles at one loop and for 4, 5 and 6 particles at two loops.

{Our approach to constructing
scattering amplitude integrands
relies on the recent study \cite{Alday:2010zy} of correlation functions of $n$ gauge-invariant composite operators $\cO$,}
\begin{equation}\label{nO}
   G_n= \vev{\cO(x_1)\cO(x_2)\dots \cO(x_n)}
\end{equation}
in $\cN=4$ super-Yang-Mills theory, in the limit where the positions of the adjacent operators become light-like separated. In \cite{Alday:2010zy} we argued that in this singular limit the correlator, divided by its Born-level expression, becomes a Wilson loop on a light-like $n-$point polygonal contour. Such Wilson loops have been shown in the past to be dual to MHV gluon scattering amplitudes, divided by their tree-level expressions \cite{am07}--\cite{Bern4}. An essential step in establishing this relation is the so-called T-duality transformation from  momenta to dual coordinates:
\begin{equation}\label{moco}
    p_i = x_i - x_{i+1} \equiv x_{i,i+1}\,, \qquad x^2_{i,i+1}=0\,, \qquad x_{i+n} \equiv x_i
\end{equation}
(with $i=1,\ldots,n$).

In the paper \cite{us} we developed further
the approach of [3] in the following way. We restricted our discussion to bilinear scalar half-BPS operators  $\cO(x)=\Tr(\phi^2(x))$. Such operators are not renormalized and thus have fixed conformal dimension equal to their canonical dimension. We  proposed to compute the $l-$loop corrections to their correlators $G_n$ by making $l$ Lagrangian insertions,
\begin{align}\label{coloop}
G^{(l)}_n(x_1, \ldots, x_n) \propto \int \prod_{i=1}^l d^D y_i\ {\cal G}^{(0)}_{n+l}(y_1,\ldots,y_l; x_1,\ldots,x_n)\,,
\end{align}
where the integrand 
\begin{equation}\label{inte}
{\cal G}^{(0)}_{n+l} = \vev{L(y_1) \ldots L(y_l)\cO(x_1)\cO(x_2)\dots \cO(x_n)}^{\rm (0)}
\end{equation}
is  the $(n+l)-$point {\it Born-level} correlator of the $n$ operators  $\cO(x_i)$ and the $l$ Lagrangian insertions.\footnote{We recall that the $\cN=4$ SYM Lagrangian, being a member of the half-BPS stress-tensor multiplet, is itself a protected operator.}

Such a procedure is well known in quantum filed theory. It has been adapted to the case of $\cN=4$ and $\cN=2$ superconformal correlators in \cite{Intriligator:1998ig,EHSSW,ESS}. A key point is that the integrand \p{inte} is a manifestly conformal rational function of the $(n+l)$ points, defined in $D=4$ dimensions. After dividing it by the Born-level expression of the $n-$point correlator $G^{(0)}_n$, we can safely go to the light-cone limit $x^2_{i,i+1} \to 0$ {and introduce the following ratio function}  (the integration points $y_i$ are kept in arbitrary positions)
\begin{align}\label{I}
 \mathcal{I}_{n+l} = \lim_{x^2_{i,i+1} \to 0}({\cal G}^{(0)}_{n+l}/G_n^{(0)})\,.
\end{align}
 Logarithmic divergences occur when we integrate  $\mathcal{I}_{n+l}$ over the insertion points. In \cite{us} we proposed to regularize  the integrals by a {\it dual infrared} dimensional regulator, i.e., to integrate with a $(D=4-2\ep)-$dimensional measure (with $\ep<0$). We compared the resulting expressions to the MHV $n-$gluon scattering amplitudes $A_n$,  computed in \cite{Anastasiou:2003kj}--\cite{Bern3} and {rewritten in the dual space \p{moco}, for arbitrary $n$ at one loop and for $n=4,5$ at two loops.} We found the following remarkable duality relation:
\begin{equation}\label{dua}
\lim_{x^2_{i,i+1} \to 0}\ln(G_n/G_n^{(0)} )= 2 \ln\left(A_n/A^{(0)}_n\right) + O(1/N_c) +O(\ep)\,.
\end{equation}
While the relation between the correlation functions and  Wilson loops established in \cite{Alday:2010zy} has a natural physical explanation, the duality \p{dua} has no obvious origin in field theory. In this aspect it resembles (and is closely related to) the duality MHV amplitudes/Wilson loops , which has so far not found its field theory explanation.\footnote{At strong coupling, a string theory mechanism for this duality was proposed in the pioneering paper \cite{am07}.}

 The reason why the duality  \p{dua} is formulated in terms of logarithms rather than simply $\lim(G_n/G_n^{(0)}) = (A_n/A^{(0)}_n)^2$, has to do with the parity-odd sector in the amplitude. Starting with $n=5$, the standard unitarity method \cite{Bern2}--\cite{Kosower:2010yk} produces a non-vanishing parity-odd contribution which however becomes $O(\ep)$ in the logarithm of the amplitude. This is crucial for matching the amplitude to the correlator $G_n$ which, as we argued in \cite{us}, is a true scalar and hence can have no pseudo-scalar sector. A related issue is the  appearance of the so-called $\mu-$terms, which are not
detected by the four-dimensional cuts \cite{Bern4, Kosower:2010yk}. Once again, these $\mu-$terms drop out from the logarithm of the amplitude. The precise reason for this property of amplitudes calculated by standard unitarity methods is not known, but it is clear that when considering dualities of the type Wilson loop/amplitude, or the newly proposed correlator/amplitude duality, we need to compare the logarithms of both objects. 

It should be pointed out, however, that the $(n+l)-$point correlator ${\cal G}^{(0)}_{n+l}$ from \p{inte}, i.e. the {\it integrand} in the left-hand side of the duality relation \p{dua}, naturally contains a parity-odd sector. We consider the correlators \p{nO} in $\cN=2$ harmonic superspace \cite{hh}, where the protected operators $\cO$ are described by Grassmann-analytic (or half-BPS) superfields (hypermultiplet bilinears), depending on half of the chiral and on half of the antichiral odd superspace variables. In contrast, the $\cN=2$ SYM Lagrangian that we insert is a {\it chiral} superfield $\mathcal{L}(x,\theta)$. In terms of component fields, this means that the inserted Lagrangian is self-dual (or chiral), 
\begin{align}
L(x) =\int d^4 \theta \, \mathcal{L}(x,\theta) =F_{\mu\nu}F^{\mu\nu} + i F_{\mu\nu}\tilde F^{\mu\nu}+\ldots = F_{\alpha\beta} F^{\alpha\beta} +\ldots\,.
\end{align}
The topological term  $i F\tilde F$ at the insertion points is responsible for the parity-odd contribution to the integrand  ${\cal G}^{(0)}_{n+l}$. At the same time, this term is a total derivative, so we know that it drops out from the {\it integral} in \p{coloop}.   Therefore, in \cite{us} we adopted the point of view that the parity-odd sector in ${\cal G}^{(0)}_{n+l}$  was a side effect of our calculation scheme, without any relevance to  the duality \p{dua}.

Soon afterwards, a new recursive procedure for generating scattering amplitudes in $\cN=4$ SYM was proposed in \cite{them}. It treats the {\it integrand} of the amplitude as a four-dimensional conformal object in momentum twistor space. Interestingly, this procedure also produces a parity-odd contribution to the integrand.\footnote{It should be mentioned that an alternative calculation of amplitudes \cite{Cachazo:2008hp} using the generalized unitarity method has already indicated that the integrand of the amplitude may have a parity-odd sector, which integrates to zero.} Most importantly, it is only by keeping the parity-even and odd sectors together that one achieves the manifest dual conformal symmetry \cite{Drummond:2006rz} of the integrand. 

These two features of the new proposal in \cite{them} -- regarding the integrand of the amplitude as a conformal object in four dimensions and the presence of a parity-odd sector integrating to zero -- are common with our construction based on correlators with Lagrangian insertions. This stimulated us to revisit our correlator calculations, this time directly comparing integrands, not integrals. {The result is reported in this note. We found strong evidence, at one and two loops, for the following remarkably simple relation:}
\begin{equation}\label{new du}
1+\sum_{l\ge 1} g^{2l} \mathcal{I}_{n+l}  = (1+g^{2l}\sum_{l\ge 1} {I}_{n+l})^2\,.
\end{equation}
{Here $\mathcal{I}_{n+l}$ is the light-cone limit of the Born-level $n-$point correlator \re{inte} with $l$ insertions defined in \re{I}}; ${I}_{n+l}$ is the twistor integrand of the $n-$particle $l-$loop MHV amplitude from \cite{them}, depending on $n$ momentum twistors for the external particles and on $l$ pairs
of auxiliary momentum twistors for the loop integrations. Up to two loops,  the duality relation \re{new du} takes the form
\begin{align}\label{explicit}
\mathcal{I}_{n+1} = 2\, {I}_{n+1}\,,\qquad
\mathcal{I}_{n+2} =  2\, {I}_{n+2}
+ ({I}_{n+1})^2\,.
\end{align}
{Note that unlike in the duality relation \p{dua}, now  there is no need to take logarithms or to neglect $O(\ep)$ terms, the relations \p{new du} and \p{explicit} are exact. Indeed, (infrared) regularization is not an issue here, it only becomes relevant if we would substitute these integrands into the (divergent) integrals defining the amplitude. We recall that when we compare $(4-2\ep)-$dimensional {\it integrals} instead of four-dimensional {\it integrands}, we have to take into account the important observation made in Refs.~\cite{Bern4,Kosower:2010yk} that parity-odd and $\mu-$terms drop out from the logarithm of the amplitude.} 

We believe that the duality between correlators with insertions and integrands of amplitudes is even more fundamental than the already known duality Wilson loops/amplitudes. {In particular, this new duality correctly captures the parity-odd terms in the integrand,  while the Wilson loop does not 
produce them at all.} In some sense the former duality should lead to the latter. But we still have to discover the underlying reason for the relation \p{new du}.

In Section 2 of this note we show in detail how the new duality works in the simplest case of $n-$point one-loop correlators and MHV amplitudes. In Section 3 we briefly describe the two-loop tests for $n=4,5,6$ which we have successfully performed.

\section{One-loop MHV amplitudes}

In this section we compare the expressions for the one-loop integrand of the $n-$particle MHV amplitude in our approach \cite{us} with the one recently proposed in Ref.~\cite{them}.

\subsection{Correlation function approach}

As explained in the Introduction, in our approach the integrand of the one-loop $n-$particle MHV amplitude is obtained
from the {\it Born-level}  correlation functions involving $n$ composite conformal primary operators $\cO(x_i)$
and one supersymmetric chiral Lagrangian $\mathcal{L}(x_0, \q_0)$ \footnote{To simplify the notation, here we denote the single insertion point  by $x_0$, while in \p{inte} they were denoted by $y_i$.}
\begin{align}\label{g41}
  &  {\cal G}^{(0)}_{n+1}(x_0; x_1, \ldots, x_n)  = \int d^4\q_0\, \vev{ \mathcal{L}(x_0, \q_0) \cO(x_1)\ldots {\cal O}(x_n)}^{\rm (0)}  + O(g^4)\,,
 \\   \notag
 &  {G}^{(0)}_{n}(x_1, \ldots, x_n)  =   \vev{ \cO(x_1)\ldots {\cal O}(x_n)}^{\rm (0)} + O(g^2)
\end{align}
{Notice that the correlator with insertion in \p{g41} is of order $g^2$ in the coupling, because the fields at the insertion point (essentially the gauge kinetic term $FF+iF\tilde F$ at this perturbative level) interact with the matter fields at the outer points.} Further, in  Eq.~\p{g41} the Lagrangian is integrated over the Grassmann variables $\q_0$ at the insertion point, but {not} over the space-time point $x_0$.  The integrand we are discussing here  is defined as the ratio
of the correlation functions $ {\cal G}^{(0)}_{n+1}/G_n^{(0)}$ in the limit where the adjacent external points $x_i$ (with $i=1,\ldots,n$) are  light-like separated, $x^2_{i,i+1} =0$ (cf. the general definition \p{I}), 
\begin{equation}\label{onlco}
\mathcal{I}_{n+1}(x_0;x_1,\ldots,x_n)=    \lim_{x^2_{i,i+1}\to 0} {{\cal G}^{(0)}_{n+1}}/{G^{(0)}_n}  \sim 
 \lim_{x^2_{i,i+1}\to 0}\int   d^4\theta_0\  \Bigl(\sum_{k=1}^n i_{k,k+1}\Bigl)^2+ O(g^2) \,.
\end{equation}
Here the building blocks $i_{k,k+1}$ correspond to the three-point functions of two fundamental scalar
fields from the Wilson operators $\cO(x_k)$ and $\cO(x_{k+1})$
and the chiral field strength at point $(x_0,\theta_0)$ from the Lagrangian density.
The explicit expression for  $i_{k,k+1}$ can be found in Ref.~\cite{us}.
Integration over the Grassman coordinate of the Lagrangian insertion $\theta_0$ yields
\begin{align} \label{ii}
\lim_{x^2_{i,i+1}\to 0}& \int d^4\theta_0 \,  \lr{\sum_{k} i_{k,k+1}}^2 = -\frac14 \sum_{k,l=1}^n \frac{[x_{0k},x_{0,k+1} ]_\alpha{}^\beta   [x_{0l},x_{0,l+1} ] _\beta{}^ \alpha}{x_{0k}^2x_{0,k+1}^2x_{0l}^2x_{0,l+1}^2} \,,
\end{align}
where
$[x,y ]_\a{}^\b \equiv x_{\a\da} \tilde y^{\da\b} - y_{\a\da} \tilde x^{\da\b} = x^\mu y^\nu [\sigma_\mu,\tilde\sigma_\nu]_\a{}^\b\,$ (see the Appendix for our spinor conventions). 
Making use of the identities \re{tra} and \re{tre}, we can easily compute the trace in \re{ii}. In doing this, we encounter the pseudo-scalar  
\begin{align}
  \ep_{\mu\nu\lambda\rho} x_{0k}^\mu x_{0,k+1}^\nu x_{0l}^\lambda x_{0,l+1}^\rho 
\equiv \epsilon(x_{0k},x_{0,k+1},x_{0l},x_{0l+1})\,.  \label{pssc}
\end{align}
Our final result is 
\begin{align}\label{I-fin}
 \mathcal{I}_{n+1}(x_0;x_1,\ldots,x_n)= \frac12\sum_{k,l=1}^n \frac{  x_{kl}^2 x_{k+1,l+1}^2  -x_{k,l+1}^2 x_{k+1,l}^2-4i \epsilon(x_{0k},x_{0,k+1},x_{0l},x_{0,l+1})}{x_{0k}^2x_{0,k+1}^2x_{0l}^2x_{0,l+1}^2}\,.
\end{align}

By construction, $ \mathcal{I}_{n+1}$ is given
by the ratio of two correlation functions of conformal operators. So, it should
transform covariantly under conformal transformations in $x-$space with zero weight at the external points and with the canonical weight 4 of the Lagrangian at the insertion point
$x_0$. {This property is manifest for the parity-even terms  in the right-hand side of \re{I-fin}, but it does not hold for the individual parity-odd terms. However, conformal invariance in the parity-odd sector gets restored in the cyclic sum of all terms (see Eqs.~\p{long} and \re{f}  below)}. Indeed, Eq.~\re{ii} represents a sum of Feynman diagrams, which are neither gauge nor conformally invariant separately, but their sum is. 

The expression for $\mathcal{I}_{n+1}$ does not have definite parity. The parity-even
part is given by a sum of terms each of which can be mapped, upon the change of variables
$p_i=x_i-x_{i+1}$, into one-loop Feynman integrands of one-mass and two-mass-easy topologies. This agrees with the well-known result for the one-loop $n-$particle MHV
amplitude \cite{Bern2}. In Ref.~\cite{us} we argued that the parity-odd part of $\mathcal{I}_{n+1}$ originates from the parity-odd part of the Lagrangian insertion $\int d^4\q_0\,  L(x_0, \q_0)$. The latter is given by a total derivative, $\epsilon_{\mu\nu\rho\lambda} F^{\mu\nu}F^{\rho\lambda} = \partial_\mu K^\mu$ and, therefore, it vanishes upon integration over the four-dimensional insertion point $x_0$. 

\subsection{Momentum twistor approach}

In the approach of Ref.~\cite{them}, the integrand of the $n-$particle amplitude is formulated in terms of $n+2$ momentum twistors defined as the following $SL(4)$ spinors
\ba
Z_i = \left(
\begin{array}{r}
  \l_{i \a}  \\
 (\tilde x_i)^{\da\a}\l_{i \a}     
\end{array}
\right)\,,\qquad Z_A = \left(
\begin{array}{r}
  \l_{A \a}  \\
 (\tilde x_0)^{\da\a}\l_{A \a}     
\end{array}
\right)\,,\qquad Z_B = \left(
\begin{array}{r}
  \l_{B\a}  \\
 (\tilde x_0)^{\da\a}\l_{B \a}     
\end{array}
\right)\,,
\end{align}
where $\a,\da=1,2$ and $i=1,\ldots,n$. Here the momentum twistors $Z_i$ 
correspond to the external particles carrying the light-like momenta
\ba\label{7}
(p_i)_{\a\da} =  (x_i - x_{i+1})_{\a\da} = \l_{i\a} \dl_{i\da}\,,
\end{align}
with $\l_{i\a}$, $\dl_{i\da}$ being (anti)chiral commuting spinors and  $x_i$ being the dual coordinates subject to the periodicity condition $x_{n+1}=x_1$. The two momentum twistors $Z_A$ and $Z_B$ are associated with the loop momentum. They depend on the same dual coordinate $x_0$ and two auxiliary spinors $\l_A$ and $\l_B$.

The integrand is expressed in terms of (dual conformal)  $SL(4)-$invariant contractions of four momentum twistors  $\vev{ABij}$ and $\vev{ijkl}$ given by
\ba
\vev{ijkl} &= \frac{1}{4!}\ep_{abcd} Z^a_i Z^b_j Z^c_k Z^d_l   = \vev{kl} \vev{i|x_{il} \tilde x_{jl}|j} + \vev{il} \vev{j|x_{jl} \tilde x_{kl}|k} + \vev{jl} \vev{k|x_{kl} \tilde x_{il}|i} \,,\label{10}
\end{align}
where $x_{ij}=x_i-x_j$, and the standard notation was used for the $SL(2)-$invariant contractions
of spinors,
\begin{align}
\vev{ij} = \l^\a_i \l_{j\a} = \ep^{\a\b}\l_{i\b} \l_{j\a}\,,\qquad
\vev{i|x\,\tilde y|j} = \l^\a_i x_{\a\da} \tilde y^{\da\b} \l_{j\b}\,.
\end{align}
The expression for  $\vev{ABij}$ is similar with $x_A=x_B=x_0$.

According to Ref.~\cite{them}, the one-loop integrand of the $n-$particle MHV amplitude  
has the following form
\begin{align}\label{I-tw}
I_{n+1} = \frac1{n} \lr{\sum_{2 < i < n} I_{n+1}^{\rm (box)}(i) + \sum_{3<i <j \le n}  I_{n+1}^{\rm (pentagon)}(i,j)} + \text{cyclic}\,,
\end{align}
where $I_{n+1}^{\rm (box)}(i)$ and $I_{n+1}^{\rm (pentagon)}(i,j)$ correspond to box and pentagon diagrams, respectively. Their explicit expressions are 
\footnote{Compared to Ref.~\cite{them}, we have inserted an additional factor $\vev{AB}^4$. This factor accounts for the different definitions of the integration measure over the loop momentum in the two approaches. In the correlation function approach this
measure takes the standard form (in four dimensions) $d^4 x_0$, while in the momentum twistor approach it is given by $d^3 Z_A d^3 Z_B = \vev{AB}^2 
\vev{A dA}\vev{B dB} d^4 x_0$.
}
\begin{align}\notag
I_{n+1}^{\rm (box)}(i) &= \frac{\vev{n123}\vev{12\,i \,i+1}\vev{AB}^4}{\vev{ABn1}\vev{AB12}\vev{AB23}\vev{AB \,i\, i+1}}\,,
\\
 I_{n+1}^{\rm (pentagon)}(i,j)&=\frac{\vev{2\,j\,i-1\,i}\vev{AB\,\overline{2\,j}}\vev{AB}^4}{\vev{AB12}\vev{AB23}\vev{AB\,i-1\,i}\vev{AB\,j-1\,j}\vev{AB\,j\,j+1}}\,,
\end{align}
with $\vev{AB\,\overline{2\,j}} \equiv
\vev{A123}\vev{B\,j-1\,j\,j+1}-\vev{A\,j-1\,j\,j+1}\vev{B123}$ and $\vev{AB} = \lambda_A^\a \lambda_{B\a}$.

Making use of the definition \re{10} and of the identity $\vev{AB\,i\,i+1} = \vev{AB} \vev{i\,i+1} x_{0,i+1}^2$, we can express both $I_{n+1}^{\rm (box)}$ and $I_{n+1}^{\rm (pentagon)}$ as rational functions of the dual $x-$variables:\begin{align}\notag
I_{n+1}^{\rm (box)}(i) &=  \frac{x_{13}^2 x_{2,i+1}^2}{x_{01}^2 x_{02}^2 x_{03}^2 x_{0,i+1}^2}\,,
\\ \label{II}
 I_{n+1}^{\rm (pentagon)}(i,j)&= \frac{f(x_0)}{x_{02}^2x_{03}^2x_{0i}^2 x_{0j}^2x_{0\, j+1}^2}\,,
\end{align}
where in the second relation the notation was introduced for the scalar function
\begin{align}\label{long}
f(x_0) \equiv  [2|\tilde x_{20}x_{0j}|j]\vev{j|x_{ji}\tilde x_{i2}|2} = \tr [x_{23}\tilde x_{30}x_{0j}\tilde x_{j,j+1} x_{j+1,i}\tilde x_{i2}] \,.
\end{align}
This function depends on five points in the dual space,
$x_2,x_3,x_j,x_{j+1},x_i$, and it has a number of interesting properties. Firstly, it is manifestly conformally covariant.
Secondly, in virtue of the relation $x_{i,i+1}^2=0$, it vanishes when $x_0$ belongs
to the lines passing through the two pairs of points $(x_2,x_3)$ and $(x_j,x_{j+1})$,
\begin{align}
f(s x_2 + \bar s x_3) = f(s  x_j +  \bar s x_{j+1}) = 0\,,
\end{align}
with $ \bar s=1-s$ and $s$ arbitrary. Finally, for $x_0\to \infty$
is scales as $f(x_0)\sim -x_0^2 \tr [x_{23} \tilde x_{j,j+1} x_{j+1,i}\tilde x_{i2}]$.
Working out the trace in \p{long}, we arrive at the following relation
\begin{align}  \notag 
f(x_0) &= \ft12 x_{02}^2 \left[ x_{3j}^2 x_{j+1,i}^2-x_{3,j+1}^2 x_{ji}^2
-4i\epsilon(x_{03},x_{0j},x_{0,j+1},x_{0i})\right]
\\ \notag
&+ \ft12x_{03}^2 \left[x_{2,j+1}^2 x_{ji}^2- x_{2j}^2 x_{j+1,i}^2
-4i\epsilon(x_{0j},x_{0,j+1},x_{0i},x_{02})\right]
\\ \notag
&+ \ft12x_{0j}^2 \left[ x_{2,j+1}^2 x_{3i}^2-x_{2i}^2 x_{3,j+1}^2
-4i\epsilon(x_{0,j+1},x_{0i},x_{02},x_{03})\right]
\\ \notag
&+\ft12 x_{0,j+1}^2 \left[ x_{3j}^2 x_{2i}^2-x_{3i}^2 x_{2j}^2
-4i\epsilon(x_{0i},x_{02},x_{03},x_{0j})\right]
\\ \label{f}
&+\ft12 x_{0i}^2 \left[ x_{2j}^2 x_{3,j+1}^2-x_{2,j+1}^2 x_{3j}^2
-4i\epsilon(x_{02},x_{03},x_{0j},x_{0,j+1})\right]\,.
\end{align}
We point out that each parity-even term in \p{f} transforms covariantly under
conformal transformations, while for the parity-odd terms this property holds only for the sum of the five terms involving cyclic shifts of indices.

Notice that unlike $I_{n+1}^{\rm (box)}(i)$, the pentagon
contribution $ I_{n+1}^{\rm (pentagon)}(i,j)$ does not have definite parity.
Substituting \re{II} and \re{f} into \re{I-tw}, we find that the parity-even part of the complete integrand $I_{n+1}$ is given by a sum of one-loop scalar integrands of various topologies (one-mass, two-mass easy, two-mass hard and three-mass). However, the two-mass-hard and three-mass integrands cancel out leading to the following result
\begin{align}\label{I2}
I_{n+1} = \frac14  \sum_{i,j=1}^n \frac{  x_{ij}^2 x_{i+1,j+1}^2  -x_{i,j+1}^2 x_{i+1,j}^2-4i \epsilon(x_{0i},x_{0,i+1},x_{0j},x_{0,j+1})}{x_{0i}^2x_{0,i+1}^2x_{0j}^2x_{0,j+1}^2}\,.
\end{align}

Finally, we compare the two expressions for the integrands, Eqs.~\re{I-fin} and \re{I2},  and observe that they coincide up to an overall {factor of two, }
\begin{align}\label{1}
\mathcal{I}_{n+1} =  2 \,I_{n+1} \,.
\end{align}
This result is in perfect agreement with the conjectured duality relation \re{explicit} at one loop.

\section{Two-loop tests}

In this section we give a brief description of the two-loop tests of the duality relation \p{explicit} that we have performed.

\subsection{$n=4$}

Let us first consider the case $n=4$. Here both the correlator and the twistor
expressions are very simple and easy to compare explicitly. The integrand of
the one- and two-loop four-point correlators were obtained by the insertion
method in \cite{ESS}. In the light-cone limit we define according to Eq.~\p{I}
\begin{eqnarray}
 {\cal I}_{4+1}(0;1,2,3,4) & = & 2 \, x^2_{13} x^2_{24} \, I_g(0;1,2,3,4) \, , \\
 {\cal I}_{4+2}(0,0';1,2,3,4) & = & x^2_{13} \, x^2_{24}
\Bigl( x_{13}^2 \, I_h(0,0';1,2,3;1,3,4) + x_{24}^2 \,
I_h(0,0';1,2,4;2,3,4)\Bigr) \nonumber \\
&+& \frac{1}{2} \Bigl(x^2_{13} \, x^2_{24} \, I_g(0;1,2,3,4)\Bigr)
\Bigl(x^2_{13} \, x^2_{24} \, I_g(0';1,2,3,4)\Bigr) \, + \,
(0 \leftrightarrow 0') \, .
\end{eqnarray}
Here we have used the notation
\begin{align}  \label{h1}
I_g(0;1,2,3,4) &= \frac{1}{x_{10}^2x_{20}^2x_{30}^2x_{40}^2}
\\  \label{h12}
I_h(0,0';1,2,3;4,5,6) &= \frac{1}{(x_{10}^2x_{20}^2x_{30}^2 )x_{0{0'}}^2( x_{4{0'}}^2
x_{5{0'}}^2x_{6{0'}}^2)}
\end{align}
for the integrands of the one- and two-loop scalar box integrals. On the other
hand, the integrand of the two-loop four-gluon amplitude expressed in terms of
dual $x$ coordinates is \cite{Anastasiou:2003kj}
\begin{equation}
I_{4+2} \, = \, \frac{1}{2} \, x_{13}^2 \, x_{24}^2 \Bigl( x_{13}^2
I_h(0,0';1,2,3;1,3,4) + x_{24}^2 h(0,0';1,2,4;2,3,4)\Bigr) \, + \,
(0 \leftrightarrow 0')\,, \label{a42}
\end{equation}
so that very clearly (see Eq.~\re{1})
\begin{equation}
{\cal I}_{4+2} \, = \, 2 \, I_{4+2} \, + \, \bigl( I_{4+1} \bigr)^2
\end{equation}
consistent with our duality proposal \p{explicit}. The momentum twistor
construction of \cite{them} yields
\begin{eqnarray}
I_{4+2} & = & \frac{1}{2} \,
\frac{\la 1234\ra\la 2341\ra\la 3412\ra \la AB\ra^4 \la CD\ra^4}{\la AB12\ra\la
 AB23\ra\la AB34\ra\la ABCD\ra\la CD12\ra\la CD34\ra\la CD41\ra} + \\
&& \frac{1}{2} \, \frac{\la 1234\ra\la 2341\ra\la 4123\ra \la AB\ra^4
\la CD\ra^4}{\la AB12\ra \la AB23\ra\la AB41\ra \la ABCD\ra\la CD23\ra\la
CD34\ra\la CD41\ra} \, + \, (AB \leftrightarrow CD)
\nonumber \, .
\end{eqnarray}
The equivalence of the last formula to \re{a42} follows
simply by substituting $\la i,i+1,j,j+1 \ra = \la i,i+1 \ra \la j,j+1
\ra x_{i,j}^2$ (here we have to define $x_i$ by $Z_i,Z_{i+1}$) and similar for
$\la A,B,i,i+1 \ra$. By definition, $\la 1234 \ra = - \la 2341 \ra$, so that
there is seemingly some ambiguity in reconstructing the numerator. However,
there is a unique choice for which the $\la i,i+1\ra$ terms cancel and the
$x$ space expression is reproduced.

\subsection{$n=5$}

Next, we move on to the considerably more complicated five-point two-loop
correlator. In \cite{us} we have described in detail how to construct its
integrand by the insertion method. Even in the
light-cone limit one finds a large sum over traces like
$\mathrm{tr}(\tilde x_{10}^{-1} x_{10'}^{-1} \tilde x_{20'}^{-1} x_{20}^{-1} \ldots)$
and thus of conformally covariant objects. The parity-even part is simple to
extract by putting $x_{i0} = x_{i0'} - x_{00'}$ and ordering the entries in each
trace. The trace of six sigma matrices drops from the sum, so that a single
application of \re{tre} is sufficient. The result for the parity-even
terms is concise, but like in the one-loop five-point integrand
${\cal I}_{5+1}$, we find a parity-odd part as well. The current procedure
yields a very large expression in this sector because
$\epsilon(x_{i0}, x_{j0},x_{k0},x_{l0})$ does not have definite conformal
properties.

To check the \emph{integrand} identity
\begin{equation}
{\cal I}_{5+2} \, = \, 2 \, I_{5+2} \, + \, \bigl( I_{5+1} \bigr)^2
\label{truelyMagic}
\end{equation}
in the parity-even sector amounts to verifying that
\begin{align}\notag
0& = \, x^2_{13} \, x^2_{24} \, \left[
 x^2_{13} \, I_h(0,0';1,2,3;1,3,4) + x^2_{24} \, I_h(0,0';1,2,4;2,3,4) - x^2_{14}
\, I_h(0,0';1,2,4;1,3,4) \right]
\\[2mm]  \notag&
+ \, x^2_{13} \, x^2_{14} \, x^2_{25} \bigl[ 2 \, I_h(0,0';1,2,5;1,3,4) \, - \,
I_h(0,0';1,2,3;1,4,5) \, - \, I_h(0,0';1,2,4;1,3,5) \bigr]
\\[2mm] & \notag
+ \, x^2_{24} \, x^2_{35} \left[  x^2_{25} \, I_p(0,0';1;2,5;3,4) \, - \,
x^2_{24} \, I_p(0,0';1;2,4;3,5) \, - \, x^2_{35} \, I_p(0,0';1;3,5;2,4) \right]
\\[2mm] \notag
& -  \ft1{4} \, x_{13}^2 \, x_{24}^2 \, I_g(0;1,2,3,4) \, x_{13}^2 \, x_{24}^2 \,
I_g(0';1,2,3,4) + \ft1{2} \,
x_{13}^2 \, x_{24}^2 \, I_g(0,1,2,3,4) \, x_{24}^2 \, x_{35}^2 \, I_g(0';2,3,4,5)
\\[2mm] \notag
& - \ft1{2} \, x_{13}^2 \, x_{24}^2 \, I_g(0;1,2,3,4) \, x_{35}^2 \, x_{14}^2 \,
I_g(0';1,3,4,5) + 2 \, I_\epsilon(0;1,2,3,4) \, I_\epsilon(0';1,2,3,4)
\\[2mm] \notag
&  + 4 \, I_\epsilon(0;1,2,3,4) \, I_\epsilon(0';2,3,4,5) + 4 \,
I_\epsilon(0;1,2,3,4) \, I_\epsilon(0';3,4,5,1)
\\[2mm] & + \text{(cyclic)} + (0 \leftrightarrow 0') \, . \label{magic}
\end{align} 
In addition to the definitions above we need here the integrands
\begin{align}\notag
I_\epsilon(0;1,2,3,4) &= \frac{\epsilon(x_{10},x_{20},x_{30},x_{40})}
{x_{10}^2 \, x_{20}^2 \, x_{30}^2 \, x_{40}^2} \, ,
\\[2mm] 
\label{penta}
I_p(0,0';1;2,3;4,5) &= \frac{x_{10'}^2}{(x_{10}^2 \,
x_{20}^2 \, x_{30}^2) \, x^2_{0{0'}} \, (x_{2{0'}}^2 \, x_{3{0'}}^2 \, x_{4{0'}}^2 \,
x_{5{0'}}^2)}
\end{align}
relating to the parity-odd one-loop terms and the integrand of the pentabox,
respectively. The product of two $I_\epsilon$ can be rewritten in terms
of $x^2_{ij}$ by the Fierz identity. The result is a very complicated
rational function, whose vanishing is not at all obvious. We have verified
\re{magic} by substituting rational values for the $x$, see below\footnote
{$I_\epsilon$ is best obtained as the trace over the four $x_{i0}$ minus
the scalar part of this.}. In \cite{us} the identity was formulated without
the $I_\epsilon$ terms, which are total derivatives. In this case it holds
only at the level of the integrals, not the integrands.

More is true, though: The principal result of this article is a numerical
check of \re{truelyMagic} and its six-point equivalent for the \emph{entire}
integrands, including both the parity-even and -odd parts.
In the notation of \cite{them}
\begin{equation}
I_{5+1} \, = \, \frac{2}{5} \frac{\la 1234\ra\la 2345\ra\la AB\ra^4}
{\la AB12\ra\la AB23\ra\la AB34\ra\la AB45 \ra} + 
\frac{\la AB\overline{25}\ra\la 2534\ra\la AB\ra^4}
{\la AB12\ra\la AB23\ra\la AB34\ra\la AB45 \ra\la AB51\ra} + (\text{cyclic})
\end{equation}
which we have demonstrated to be equal to ${\cal I}_{5+1}/2$ in Section 2,
and
\begin{eqnarray}
I_{5+2} & = & \frac{1}{2} \, \frac{\la 1234\ra \la 2345\ra\la 5123\ra\la
AB\ra^4\la CD\ra^4}
{\la AB12\ra\la AB23\ra\la AB51\ra\la ABCD\ra\la CD23\ra\la CD34\ra
   \la CD45\ra} \\
& + & \frac{1}{2} \, \frac{\la 1345\ra\la 3451\ra\la AB\overline{13}\ra\la
AB\ra^4\la CD\ra^4}
{\la AB12\ra\la AB23\ra\la AB34\ra\la AB51\ra\la ABCD\ra\la CD34\ra\la CD45\ra
   \la CD51\ra} \nonumber \\
& + & (AB \leftrightarrow CD) + (\text{cyclic}) \, . \phantom{\Bigr(} \nonumber
\end{eqnarray} 

To test \re{truelyMagic} by substituting numbers\footnote{We are grateful to
Nima Arkani-Hamed for suggesting this procedure to us.} is simple: We have
chosen $Z_1,\ldots,Z_5;Z_A,Z_B;Z_C,Z_D$ as vectors with four  
components of the form $n_1 + i n_2$,  with $n_1$ and $n_2$  being randomly generated
integers. Since
\begin{equation}
\la AB\overline{ij}\ra \, = \, \la A,i-1,i,i+1\ra \la B,j-1,j,j+1\ra
\, - \, \la A,j-1,j,j+1\ra \la B,i-1,i,i+1\ra
\end{equation}
all momentum twistor structures can simply be evaluated by putting the four
twistors in $\la ijkl\ra$ into a matrix and taking its determinant.
In order to evaluate our expressions we treat $Z \, = \, (\lambda_\alpha,\mu_\da)$
as a row vector. Let
\begin{equation}
M^{\dot\alpha\alpha}(\lambda,\mu) \, = \,
\left(
\begin{array}{r}
  \mu_2  \\
  - \mu_1   
\end{array}
\right)
\times
\left(
\begin{array}{rr}
  \lambda_2 & - \lambda_1   
\end{array}
\right) \, , \qquad
\la \lambda \rho\ra = \lambda_2\rho_1- \lambda_1\rho_2 \, .
\end{equation}
With these definitions we have
\begin{equation}
\tilde x_i^{\dot\alpha \alpha} \, = \,
\frac{M^{\dot\alpha \alpha}(\lambda_i,\mu_{i+1}) -
M^{\dot\alpha \alpha}(\lambda_{i+1},\mu_i)}{\la\lambda_i \lambda_{i+1}\ra}
\end{equation}
and the same with lower indices. The square of a vector can 
be obtained as $x_i^2=\frac12 \tilde x_i^{\dot\alpha \alpha}  {x_i}_{\alpha\dot\alpha }$. The evaluation of the aforementioned trace covariants
is thus reduced to matrix multiplication.

The on-shell conditions $x_{i,i+1}^2 = 0$ are solved by construction, so that
all components of the twistors are unconstrained. We have successfully run
the check for hundreds of points composed of complex integers with
real and imaginary parts in the range $\{-100,\ldots,100\}$.
{\sl Mathematica} can then do exact computations so that any disagreement would
immediately be noticed.


\subsection{$n=6$}

The insertion method with ${\cal N} = 2$ superfields can easily be applied to
the six-point correlator, too. In the following paragraph we briefly
mention a few technical points that change with respect to the discussion of the five-point case presented in Appendix A.4 of \cite{us}. We refer the interested
reader to the explanations given there.

The correlator carries charge 2 at all six outer points and its Grassmann
expansion starts at $\theta^8$ as before. This allows for the structures 
\begin{eqnarray}
A & : & (\q_1)^2 (\q_2)^2 (\q_3)^2 (\q_4)^2 (56)^2 \, f_A(x) \, \nonumber \\
B & : & (\q_1)^2 (\q_2)^2 (\q_3)^2 \q_4^\alpha \q_5^\beta (46)(56) \, f_{B \,
\alpha \beta}(x) \\
C & : & (\q_1)^2 (\q_2)^2 \q_1^\alpha \q_2^\beta \q_3^\gamma \q_4^\delta
\bigl( (34)(56) \, f_{C \,\alpha \beta \gamma \delta}(x) \, + \, (35)(46) \,
g_{C \,\alpha \beta \gamma \delta}(x) \bigr) \nonumber
\end{eqnarray}
and their point permutations.  In the third case we need not write a
coefficient for the harmonic structure $(36)(45)\equiv(u_3 u_6)(u_4 u_5)$ because it is related
to the other two by the cyclic identity.   Here $f$ and $g$ are coefficient functions
which we can once again determine  by identifications of the harmonic variables $u$.  In case A we may identify all harmonics with $u_5$ barring for $u_6$. Note that
these terms do not automatically drop as they did for the five-point correlator.
In case B we can essentially do the same; we can read off the coefficient
$f_B$ upon identifying all harmonics with, say, $u_4$ barring for $u_6$. In
case C we must consider two different projections in order to identify the two
distinct coefficient functions. First, we identify $u_1 = u_2 = u_3 = u_5$ and
$u_4 = u_6$ which sends the $g_C$ term to zero, so that $f_C$ can be read off.
Next we identify $u_1 = u_2 = u_3 = u_4$ and $u_5 = u_6$ which projects out
$f_C$ and allows to read off $g_C$. As before this allows to sidestep the use
of the cyclic identity, which would be very heavy given the number of points.
Second, and most importantly, in all sets of identifications that we have
chosen, the ``TT-block'' (two insertions connected to one matter line)
is suppressed. The rest of the calculation is strictly analoguous to the
five-point case, notably the light-cone limit selects the graphs inscribed
into the planar hexagon $123456$. The relevant set of graphs is listed in
Figure 1.

Although the reconstructed integrand is still larger than in the five-point
case, it contains the same trace structures as before whereby the
numerical evaluation could be done by the same set of routines.
The integrand satisfies the six-point version of Eq.~\re{truelyMagic}.

\begin{figure}[th]
\begin{center}
\includegraphics[width = 0.75 \textwidth]{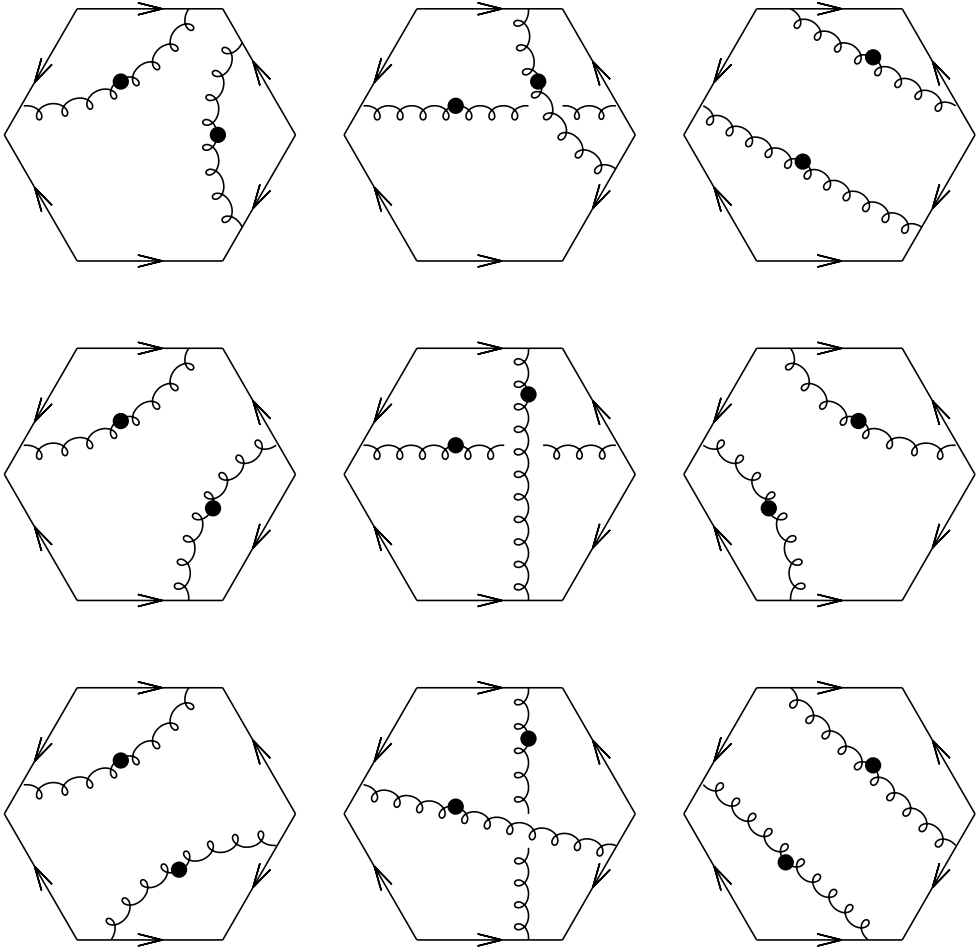}
\vskip 0.2 cm
\caption{Graphs contributing to the six-point correlator
on the light cone}
\end{center}
\vskip 0.5 cm
\end{figure}

\section{Conclusions}

In summary, we have presented convincing evidence  that two different approaches, one
based on the analysis of the correlation functions with Lagrangian
insertions \cite{Alday:2010zy,us}
and the other on the recursive BCFW construction in momentum twistor space \cite{them,Boels:2010nw},
lead to identical expressions for the four-dimensional integrands
of MHV amplitudes. The next obvious step would be to
extend this correspondence to non-MHV amplitudes and, eventually,
identify a dual object analogous to the light-like Wilson loop which could describe
generic superamplitudes in planar $\mathcal{N}=4$ SYM
theory~\cite{Coming papers1,Coming papers,Belitsky:2011zm}. Another, major challenge is to understand {\it why} this duality works. What makes the two integrands identical? Are there some hidden symmetries or some recursive structure of the correlator, which fix it to a unique form? We hope to find the answers to these questions soon.

\section*{Acknowledgments}

During this work we have profited a lot from discussions with Nima Arkani-Hamed, Simon Caron-Huot and Jake Bourjaily who shared with us their insight in the twistor formulation of scattering amplitudes.  Special thanks are due to Jake Bourjaily who provided us with a {\it Mathematica} code for generating twistor integrands. {ES is grateful to Juan Maldacena  and the Institute for Advanced Study, and to George Zoupanos and the Corfu Summer Institute  for warm hospitality at various stages of this work.} We would like to thank Johannes Henn and David Kosower for useful 
discussions. This work was supported in part by the French Agence Nationale de la Recherche under grant
ANR-06-BLAN-0142. BE acknowledges support by STFC under the rolling grant ST/G000433/1.

\appendix

\section{Spinor conventions}

We use the Pauli matrices $\vec\sigma=(\sigma_1,\sigma_2,\sigma_3)$ to 
define two sets of Minkowski space sigma matrices $\sigma^\mu=(1,\vec \sigma)$ and $\tilde\sigma^\mu=(1,-\vec \sigma)$.
They are related to each other by raising (or lowering) their two-component indices with the help of the Levi-Civita symbol,
\begin{align}
 (\tilde\sigma^\mu)^{\da\a} = \ep^{\da\db}\ep^{\a\b}(\sigma^\mu)_{\b\db}\,, \qquad \ep^{12}=\ep^{\dot 1 \dot 2} = -1\,,
\end{align} 
 satisfy the anticommutator relation 
 \ba
\sigma^\mu\tilde\sigma^\nu  + \sigma^\nu\tilde\sigma^\mu = 2 g^{\mu\nu}\mathbb{I}\,, \qquad g^{\mu\nu} = (+---) \,,
\end{align}
and have  the trace
\ba\label{tra}
\tr{\sigma^\mu\tilde\sigma^\nu} \equiv (\sigma^\mu)_{\a\da}(\tilde\sigma^\nu)^{\da\a} = 2 g^{\mu\nu} \,.
\end{align}
The product of three matrices is decomposed into a linear combination of single matrices  
\ba\label{tre}
\sigma^\mu\tilde\sigma^\nu\sigma^\lambda = g^{\mu\nu}\sigma^\lambda  -g^{\mu\lambda}\sigma^\nu  + g^{\lambda\nu}\sigma^\mu + i \ep^{\mu\nu\lambda\rho} \sigma_\rho\,, \qquad \ep^{0123}=-1\,.   
\end{align}
Using \p{tra} and \p{tre}, one can compute the trace of the product of any even number of sigma matrices.
Minkowski four-vectors $x_\mu = (x_0, x_1, x_2, x_3)$ have the following forms 
in matrix notation
\ba
&x_{\a\da} = x_\mu (\sigma^\mu)_{\a\da} \,, \qquad
 \tilde x^{\da\a} = x_\mu (\tilde\sigma^\mu)^{\da\a} \,.
\end{align}
The square of the vector is given by the determinant of the matrix:
$$
x^2=x_\mu x^\mu = x^2_0-x^2_1-x^2_2-x^2_3 = \det(x) = \det(\tilde x)\,.
$$
Matrices are multiplied by alternating $\sigma$ and $\tilde\sigma$, e.g.,
\ba
x_{\a\da}\tilde x^{\da\b} = x^2\, \delta_\a^\b \,.
\end{align}


\end{document}